\documentclass[12pt]{article}
\usepackage[english]{babel}
\usepackage{amsmath}
\usepackage{amssymb}
\usepackage{bbm}
\usepackage{color}
\usepackage{graphicx}
\usepackage{epsfig}
\usepackage[titletoc,toc,title]{appendix}
\Roman{section}
\newcommand{\naw}[1]{\left(#1\right)}
\newcommand{\ket}[1]{\left|#1\right>}
\newcommand{\bra}[1]{\left<#1\right|}
\newcommand{\av}[1]{\left<#1\right>}

\newcommand{\modu}[1]{\left|#1\right|}
\newcommand{\poisson}[1]{\left\{#1\right\}}

\begin{document}

\begin{center}
\textsc{\Large{Three-player conflicting interest games and nonlocality}}
\newline

\large{Katarzyna Bolonek-Laso\'n}\footnote{kbolonek@uni.lodz.pl}\\ 
\emph{\normalsize{Faculty of Economics and Sociology, Department of Statistical Methods, \\University of Lodz,
41/43 Rewolucji 1905 St., 90-214 Lodz,  Poland.}}\\

\end{center}
\begin{abstract}
We outline the general construction of three-player games with incomplete information which fulfill the following conditions: (\textit{i}) symmetry with respect to the permutations of players; (\textit{ii}) the existence of an upper bound for total payoff resulting from Bell inequalities; (\textit{iii}) the existence of both fair and unfair Nash equilibria saturating this bound. Conditions (\textit{i})$\div$(\textit{iii}) imply that we are dealing with conflicting interest games. An explicit example of such a game is given. A quantum counterpart of this game is considered. It is obtained by keeping the same utilities but replacing classical advisor by a quantum one. It is shown that the quantum game possesses only fair equilibria with strictly higher payoffs than in the classical case. This implies that quantum nonlocality can be used to resolve the conflict between the players.
\end{abstract}

\section{Introduction}
One of the most important features of quantum theory is the nonlocality - the existence of correlations that cannot be explained in the framework of any local realistic theory. In particular, the correlations admitted by the latter must satisfy certain set of inequalities which can be violated on the quantum level \cite{Bell}.

Violation of Bell inequalities has been confirmed experimentally \cite{Aspect}. Nonlocality inherent to quantum physics appears to be useful in practice, in particular for information processing (see \cite{Ekert}, \cite{Acin}, \cite{Buhrman} and numerous other references). Bell inequalities can be also discussed in the context of game theory. One can pose the question how the properties of the game are modified due to the existence of nonlocal correlations between the players. The first attempts to construct the games based on quantum mechanical correlations concerned those of complete information \cite{Meyer}$\div$\cite{Flitney}. It appeared that the quantum versions of classical games offer additional strategies which allow to resolve dilemmas that occur in classical games (for example, the Prisoner's Dilemma). It has been also shown that quantum games can be realized experimentally \cite{Du}, \cite{Prevedel}.

On the other hand the conclusion that the advantages of quantum counterparts of classical games with complete information result from the specific properties of quantum correlations has been much debated and criticized \cite{Benjamin}, \cite{Enk}. In order to make the relation between quantum nonlocality and the advantages of quantum strategies  more transparent  the quantum versions of games with incomplete information \cite{Harsanyi} have been proposed \cite{Cheon1}. In this way the connection has been established between the Bell theorem and the Bayesian games. In order to understand it let us note that, as it is nicely explained  by Fine \cite{Fine}, \cite{Fine1} (see also \cite{Halliwell}, \cite{Halliwell1}), the violation of Bell inequalities is directly related to the existence of noncommuting observables. Now, the unknown elements of the game with incomplete information are represented by the concept of the player type. On the quantum level the player types are, in turn,  represented by different, in general noncommuting, observables. This leads to the violation of Bell inequalities. If the payoff functions of the players are related to Bell expressions, the players sharing nonlocal resources can outperform the ones having access to classical resources only.

In the particular example proposed by Cheon and Iqbal \cite{Cheon1}, which is a mixture of Battle of Sexes and Chicken games, the bound on classical payoffs is related to Cereceda inequalities \cite{Cereceda1}. The ideas of Cheon and Iqbal were further developed in the papers \cite{Iqbal}, \cite{Flitney1}, \cite{Iqbal1}, \cite{Hill}. Quite recently Brunner and Linden \cite{Brunner} considered the more general situation where nonlocal resources provide an advantage over any classical strategy because the bounds on some combinations of payoff functions follow from Bell inequalities.

The examples of games presented in \cite{Brunner} are the games of common interest. Pappa et al. \cite{Pappa} gave an example of conflicting interest game where quantum mechanics also offers an advantage over its classical counterpart. The game they consider is a two-player one obtained as a combination of the Battle of Sexes and CHSH games. It is symmetric with respect to the permutation of players although this property is somewhat hidden due to the specific numbering of strategies and types used by the authors. An important point is that, in the classical version of the game, the total payoff (the sum of payoffs of both players) is bounded from above due to the Bell inequality. This implies that it is a conflicting interest game provided there exist unfair equilibria saturating the bound resulting from Bell inequality. On the other hand, on the quantum level all equilibria are the fair ones because the payoff functions become equal. Moreover, there exist fair quantum equilibria where the parties have strictly higher payoffs than for any classical fair equilibrium.

Further examples of conflicting interest games where quantum mechanics offers an advantage were given by Situ \cite{Situ}. Moreover, by a slight modification of the payoff functions proposed in \cite{Pappa} Roy et.\,al \cite{Roy} gave the examples of games where quantum strategies can outperform even the unfair classical equilibrium strategies. Depending on the values of additional parameters entering the payoff functions their games have only fair equilibria, both fair and unfair equilibria or only unfair ones.

In final instance, all physics behind any example of quantum game with incomplete information is related to quantum nonlocality manifesting in violation of some Bell inequalities. The advantages of quantum strategies are the consequence of quantum entanglement built into the game. 

In the multipartite case the structure of nonlocal correlations is richer and less understood \cite{Brunner1}. There exist different definitions of nonlocality which refine the bipartite definition. Therefore, it is advantegous to consider the three- or multiparty generalizations of quantum Bayesian games. One example of three-party (three-player) game has been provided by Situ et al. \cite{Situ1}. It is based on Svetlichny inequality \cite{Svetlichny} and allows to analyse the advantages of the game based on fully quantum correlations over the one where the correlations can be reduced to the mixtures of two-player ones related locally to the third player. 

In the present paper we consider the three-player counterpart of the game considered in Ref. \cite{Pappa}. In Sec.\,II we outline the general construction of the three-player games with incomplete information possessing the upper bound for total payoff following from Bell inequalities and both fair and unfair equilibria saturating this bound. As we have mentioned above such games are automatically conflicting interest games. In Sec.\,III we provide an explicit example of such a game. As a next step we consider in Sec.\,IV a quantum counterpart of our game, i.e. we keep the utilities intact but replace the classical advisor by a quantum one. It appears that the quantum game possesses only fair equilibria and the corresponding payoffs are strictly higher than the classical ones. We show that the nonlocality inherent in quantum mechanics plays the twofold role: it raises, due to violation of Bell inequalities, the payoffs corresponding to fair equilibria and excludes the unfair ones.

To conclude this section let us note that the game-theoretic language has much wider range of applications and is a very convenient tool for describing the peculiar properties of quantum correlations. It can be used, for example, to study the entanglement in spin systems \cite{Miszczak}, \cite{Ozaydin} or decoherence phenomena \cite{Gawron}, \cite{Gawron1}, \cite{Dajka}.

\section{Three-player games}

We define a three-player Bayesian game following the analogous discussion of two-player case by Pappa et al. \cite{Pappa} (see also \cite{Brunner}). There are three players, Alice (A), Bob (B) and Charlie (C); each player acquires a type $x_i$, $i\in\poisson{A,B,C}$, $x_i\in\poisson{0,1}$, according to the probability distribution $P(\underline{x})\equiv P\naw{x_A,x_B,x_C}$. They decide on their actions $y_i$, $y_i\in\poisson{0,1}$, according to a chosen strategy. The average payoff of each player reads
\begin{equation}
F_i=\sum_{\naw{x,y}}P\naw{\underline{x}}p\naw{\underline{y}|\underline{x}}u_i\naw{\underline{x},\underline{y}}
\end{equation}
where $p\naw{\underline{y}|\underline{x}}\equiv p\naw{y_A,y_B,y_C|x_A,x_B,x_C}$ is the probability the players choose actions $\underline{y}\equiv\naw{y_A,y_B,y_C}$ given their types were $\underline{x}\equiv\naw{x_A,x_B,x_C}$; $u_i\naw{\underline{x},\underline{y}}$ are the utility functions determining the gains of players depending on their types and actions. 
The properties of the game are determined by the form of utility functions and the restrictions imposed on the probabilities $p\naw{\underline{y}|\underline{x}}$. 

In order to set the question of fair and unfair equilibria in the proper framework we consider the games symmetric with respect to the permutations of players. This implies the relations (which hold up to a possible renumbering of types and/or strategies)
\begin{equation}
\begin{split}
& u_A\naw{x_A,x_B,x_C,y_A,y_B,y_C}=u_B\naw{x_B,x_A,x_C,y_B,y_A,y_C}\\
& u_C\naw{x_A,x_B,x_C,y_A,y_B,y_C}=u_C\naw{x_B,x_A,x_C,y_B,y_A,y_C}
\end{split}\label{a}
\end{equation}
together with the similar relations obtained by choosing the remaining two pairs of players. As far as the probabilities $p\naw{\underline{y}|\underline{x}}$  are concerned we assume they obey the no-signalling conditions \cite{Brunner1}
\begin{equation}
\sum_{y_C}p\naw{y_A,y_B,y_C|x_A,x_B,x_C}=\sum_{y_C}p\naw{y_A,y_B,y_C|x_A,x_B,x_C'\label{aa}}
\end{equation}
and similar conditions for the remaining two players.
Apart from eq. (\ref{aa}) we have the normalization conditions  
\begin{equation}
\sum_{\underline{y}}p\naw{\underline{y}|\underline{x}}=1 \quad \text{for all} \quad \underline{x}.
\end{equation}
Given no-signalling condition we consider two types of probability distributions:
\begin{itemize}
\item[(i)] \underline{the classical case}: one assumes further constraints on $p\naw{\underline{y}|\underline{x}}$ in form of Bell inequalities \cite{Hardy}, \cite{Cereceda2}, \cite{Brunner1}. According to Fine \cite{Fine}, \cite{Fine1}, \cite{Halliwell}, \cite{Halliwell1} this leads to the hidden variables representation of the relevant probabilities (actually, Fine's theorem concerns two-parties case but we assume it holds for three-parties as well):
\begin{equation}
p\naw{y_A,y_B,y_C|x_A,x_B,x_C}=\int \text{d}\lambda \rho(\lambda)p_A\naw{y_A|x_A,\lambda}\cdot p_B\naw{y_B|x_B,\lambda}\cdot p_C\naw{y_C|x_C,\lambda}\label{a4}
\end{equation}
$\lambda$ being a set of hidden variables distributed with probability density $\rho\naw{\lambda}$. Note that since there are only two possible actions per player it is sufficient to consider only hidden variables providing three bits so that $\lambda\equiv\naw{\lambda_A,\lambda_B,\lambda_C}$, $p_A\naw{y_A|x_A,\lambda}=p_A\naw{y_A|x_A,\lambda_A}$ etc.\\
In game-theoretic language one says that the players receive advice from a classical source that is independent of the inputs $\underline{x}$; $\rho\naw{\lambda}$ can be viewed as characterizing an advisor.
In particular, if the strategies of the players are uniquely determined by their types and advices they received (deterministic hidden variable model),
\begin{equation}
y_A=c_A\naw{x_A,\lambda},\quad \text{etc.}
\end{equation}
one finds
\begin{equation}
F_i=\sum_{\underline{x}}P\naw{\underline{x}}\int\text{d}\lambda \rho(\lambda)u_i\naw{x_A,x_B,x_C,c_A\naw{x_A,\lambda},c_B\naw{x_B,\lambda},c_C\naw{x_C,\lambda}}
\end{equation}
On the other hand if the players are insensitive to the advisor suggestions, $p_A\naw{y_A|x_A,\lambda}=p_A\naw{y_A|x_A}$, etc., the probability factorizes
\begin{equation}
p\naw{y_A,y_B,y_C|x_A,x_B,x_C}=p_A\naw{y_A|x_A}p_B\naw{y_B|x_B}p_C\naw{y_C|x_C}
\end{equation}
\item[(ii)] \underline{the quantum case}: quantum probabilities (a quantum source/advisor) are defined by the choice of tripartite density matrix $\rho$ (which characterizes an advisor) and the choice of three pairs of observables $A_x$, $B_x$ and $C_x$, $x=0,1$, acting in twodimensional Hilbert spaces of individual players and admitting the spectral decompositions
\begin{equation}
A_x=1\cdot A_x^1+\naw{-1}\cdot A_x^0, \quad \mathbbm{1}=A_x^1+A_x^0,\quad \text{etc.}
\end{equation}
with $A_x^y$, etc., being the corresponding projectors. The resulting payoffs read
\begin{equation}
F_i=\sum_{\underline{x},\underline{y}}P\naw{\underline{x}}\text{Tr}\naw{\rho\naw{A_{x_A}^{y_A}\otimes B_{x_B}^{y_B}\otimes C_{x_C}^{y_C}}}u_i\naw{\underline{x},\underline{y}}.\label{b2}
\end{equation}
\end{itemize}
Note that the general form of our quantum variables reads
\begin{equation}
A_x=\vec{n}_x^{(A)}\cdot\vec{\sigma}
\end{equation}
where $\vec{\sigma}$ are Pauli matrices while $\vec{n}_{0,1}^A$- the unit vectors,\\
 $\vec{n}_x^{(A)}=\naw{\sin\theta_x^A\cos\varphi_x^A,\sin\theta_x^A\sin\varphi_x^A,\cos\theta_x^A}$; similar formulae are valid for $B$ and $C$.
\\

In principle, we could also consider superquantum no-signalling distributions \cite{Popescu}; however, we shall not dwell on this question.

In what follows we assume that the distribution of the player types is uniform,
\begin{equation}
P\naw{\underline{x}}=\frac{1}{8}\quad \text{for all}\quad \underline{x}=\naw{x_A,x_B,x_C}. \label{ca}
\end{equation}
In order to construct the examples of games with conflicting interests which possess fair quantum equilibria with higher payoffs than those corresponding to classical equilibria we start with the utility functions $u_i\naw{\underline{x},\underline{y}}$, $i\in\poisson{A,B,C}$. We demand they obey the symmetry conditions (\ref{a}). Moreover, we assume that the sum of payoffs $F_A+F_B+F_C$ is expressible in terms of the expression(s) entering the Bell inequality(ies). The relevant Bell inequality reads \cite{Hardy}, \cite{Cereceda2}, \cite{Brunner1}
\begin{equation}
\modu{\av{A_0B_1C_1}+\av{A_1B_0C_1}+\av{A_1B_1C_0}-\av{A_0B_0C_0}}\leq 2\label{a2}
\end{equation}
where $A_x$, $B_x$ and $C_x$ acquire the values $\pm 1$. Rewritting the above inequality in terms of relevant probabilities and comparying the resulting expression with $F_A+F_B+F_C$ one finds the conditions on utility functions. However, there is an important difference between two- and three-players cases. In the latter one the resulting equations are more stringent and imply that the utility functions lead to a trivial game. This can be cured as follows. Note that the properties of a game (i.e. the structure of its Nash equilibria) are invariant under the transformations
\begin{equation}
u_i\naw{\underline{x},\underline{y}}\rightarrow \alpha u_i\naw{\underline{x},\underline{y}}+\beta\label{a1}
\end{equation}
with arbitrary $\alpha$ and $\beta$. Therefore, if the constraints on $u_i's$ are not invariant under (\ref{a1}) their solutions must be so special that they lead to a trivial game.

However, note that we have eight Bell inequalities at our disposal. In fact, the remaining ones are obtained from (\ref{a2}) by making the replacement $0\leftrightarrow 1$ for one, two and three players. In particular, in the latter case we arrive at the inequality
\begin{equation}
\modu{\av{A_1B_0C_0}+\av{A_0B_1C_0}+\av{A_0B_0C_1}-\av{A_1B_1C_1}}\leq 2.\label{a3}
\end{equation}
By demanding that $F_A+F_B+F_C$ is expressible in terms of the linear combination (actually, the difference) of expressions entering  (\ref{a2}) and (\ref{a3}) one finds much more reasonable conditions on utility functions (in particular, they are invariant under the transformations (\ref{a1})).

The symmetry conditions (\ref{a}) and the one imposed on $F_A+F_B+F_C$ allow us to express the utilities $u_i\naw{\underline{x},\underline{y}}$ in terms of a number of independent parameters.\\
As a next step we select some set of strategies as the candidates for nonfair classical equilibria. Additionally, we demand that, for these equilibria, the sum $F_A+F_B+F_C$ saturates the uper bound following from Bell inequalities. If this is the case we can take for granted that, for any fair equilibrium, at least one player will gain smaller payoff than for the unfair one.
The resulting general conditions (derived with the help of MATHEMATICA) are too complicated to present them here explicitly. Instead, we give an example of a game sharing the properties discussed above.

\section{The example of three-player game}
 The utilities in our example are presented in Table \ref{t1}.

\begin{table}
\caption{The utilities of players}
\begin{footnotesize}
 $u_A\naw{\underline{x},\underline{y}}$:
\begin{tabular}{|c|c|c|c|c|c|}\cline{3-6}
\multicolumn{2}{c}{}& \multicolumn{2}{|c|}{$y_C=0$}&\multicolumn{2}{|c|}{$y_C=1$}\\
\cline{3-6}
\multicolumn{2}{c|}{} & $y_B=0$ & $y_B=1$ & $y_B=0$ & $y_B=1$\\
\hline
$x_C=0$ & $x_B=0$ & $\left[\begin{array}{cc} 2 & 0\\ 2 & 1\end{array}\right]$ &  $\left[\begin{array}{cc} \frac{3}{2} & 1\\ 0 & 2\end{array}\right] $ &  $\left[\begin{array}{cc} \frac{3}{2} & 1\\ 0 & 2\end{array}\right]$  & $\left[\begin{array}{cc} 4 & 1\\ 4 & \frac{19}{3}\end{array}\right] $ \\ 
\cline{2-6}
 & $x_B=1$ &  $\left[\begin{array}{cc} 0 & -1\\ -1 & 1\end{array}\right]$  & $ \left[\begin{array}{cc} -\frac{1}{2} & 2\\ 1 & 0\end{array}\right] $  &  $\left[\begin{array}{cc} 1 & -1\\ \frac{1}{2} & 0\end{array}\right] $ & $ \left[\begin{array}{cc} -2 & -\frac{19}{6}\\ -1 & -\frac{1}{2}\end{array}\right]  $\\
  \hline
$x_C=1$ & $x_B=0$ &  $\left[\begin{array}{cc} 0 & -1\\ -1 & 1\end{array}\right]$   & $\left[\begin{array}{cc} 1& -1\\ \frac{1}{2} & 0\end{array}\right] $  &  $\left[\begin{array}{cc} -\frac{1}{2} & 2\\ 1 & 0\end{array}\right] $ &  $\left[\begin{array}{cc} -2 & -\frac{19}{6}\\ -1 & -\frac{1}{2}\end{array}\right] $\\
\cline{2-6}
 & $x_B=1$ &  $\left[\begin{array}{cc} 2 & 2\\ 0 & -2\end{array}\right] $ & $\left[\begin{array}{cc} 1 & 1\\ 2 & \frac{1}{2}\end{array}\right]$  & $ \left[\begin{array}{cc} 1 & 1\\ 2 & \frac{1}{2}\end{array}\right]$  &  $\left[\begin{array}{cc} 0 & 4\\ -1 & \frac{2}{3} \end{array}\right]$ \\
  \hline
\end{tabular}\\
$ u_B\naw{\underline{x},\underline{y}}$:
\begin{tabular}{|c|c|c|c|c|c|}\cline{3-6}
\multicolumn{2}{c}{}& \multicolumn{2}{|c|}{$y_C=0$}&\multicolumn{2}{|c|}{$y_C=1$}\\
\cline{3-6}
\multicolumn{2}{c|}{} & $y_B=0$ & $y_B=1$ & $y_B=0$ & $y_B=1$\\
\hline
$x_C=0$ & $x_B=0$ & $\left[\begin{array}{cc} 2 & \frac{3}{2}\\ 0 & -\frac{1}{2}\end{array}\right]$ &  $\left[\begin{array}{cc} 0 & 1\\ -1 & 2\end{array}\right]$  &  $\left[\begin{array}{cc} \frac{3}{2} &4\\ 1 & -2\end{array}\right]$  & $\left[\begin{array}{cc} 1 & 1\\ -1 & -\frac{19}{6}\end{array}\right]$  \\ 
\cline{2-6}
 & $x_B=1$ &  $\left[\begin{array}{cc} 2 & 0\\ -1 & 1\end{array}\right]$  &  $\left[\begin{array}{cc} 1 & 2\\ 1 & 0\end{array}\right]$   &  $\left[\begin{array}{cc} 0 & 4\\ \frac{1}{2} & -1\end{array}\right]$  & $\left[\begin{array}{cc} 2 & \frac{19}{3}\\ 0 & -\frac{1}{2}\end{array}\right]$  \\
  \hline
$x_C=1$ & $x_B=0$ &  $\left[\begin{array}{cc} 0 & 1\\ 2 & 1\end{array}\right]$   & $\left[\begin{array}{cc} -1 & -1\\ 2 & 1\end{array}\right]$   &  $\left[\begin{array}{cc} -\frac{1}{2} & -2\\ 1 & 0\end{array}\right]$  &  $\left[\begin{array}{cc} 2 & -\frac{19}{6}\\ 1 & 4\end{array}\right]$ \\
\cline{2-6}
 & $x_B=1$ &  $\left[\begin{array}{cc}-1 & \frac{1}{2}\\ 0 & 2\end{array}\right]$  & $\left[\begin{array}{cc} 1 & 0\\ -2 & \frac{1}{2}\end{array}\right]$   &  $\left[\begin{array}{cc} 1 & -1\\ 2 & -1\end{array}\right]$  &  $\left[\begin{array}{cc} 0 & -\frac{1}{2}\\ \frac{1}{2} & \frac{2}{3}\end{array}\right]$ \\
  \hline
\end{tabular}
\\
$ u_C\naw{\underline{x},\underline{y}}$:
\begin{tabular}{|c|c|c|c|c|c|}\cline{3-6}
\multicolumn{2}{c}{}& \multicolumn{2}{|c|}{$y_C=0$}&\multicolumn{2}{|c|}{$y_C=1$}\\
\cline{3-6}
\multicolumn{2}{c|}{} & $y_B=0$ & $y_B=1$ & $y_B=0$ & $y_B=1$\\
\hline
$x_C=0$ & $x_B=0$ & $\left[\begin{array}{cc} 2 & \frac{3}{2}\\ 0 & -\frac{1}{2}\end{array}\right]$ &  $\left[\begin{array}{cc} \frac{3}{2} & 4\\ 1 & -2\end{array}\right]$  &  $\left[\begin{array}{cc} 0 & 1\\ -1 & 2\end{array}\right]$  & $\left[\begin{array}{cc} 1 & 1\\ -1 & -\frac{19}{6}\end{array}\right]$  \\ 
\cline{2-6}
 & $x_B=1$ &  $\left[\begin{array}{cc} 0 & 1\\ 2 & 1\end{array}\right]$  &  $\left[\begin{array}{cc} -\frac{1}{2} & -2\\ 1 & 0\end{array}\right]$   &  $\left[\begin{array}{cc} -1 & -1\\ 2 & 1\end{array}\right]$  & $\left[\begin{array}{cc} 2 & -\frac{19}{6}\\ 1 & 4\end{array}\right]$  \\
  \hline
$x_C=1$ & $x_B=0$ &  $\left[\begin{array}{cc} 2 & 0\\ -1 & 1\end{array}\right]$   & $\left[\begin{array}{cc} 0 & 4\\ \frac{1}{2} & -1\end{array}\right]$   &  $\left[\begin{array}{cc} 1 & 2\\ 1 & 0\end{array}\right]$  &  $\left[\begin{array}{cc} 2 & \frac{19}{3}\\ 0 & -\frac{1}{2}\end{array}\right]$ \\
\cline{2-6}
 & $x_B=1$ &  $\left[\begin{array}{cc}-1 & \frac{1}{2}\\ 0 & 2\end{array}\right]$  & $\left[\begin{array}{cc} 1 & -1\\ 2 & -1\end{array}\right]$   &  $\left[\begin{array}{cc} 1 & 0\\ -2 &\frac{1}{2}\end{array}\right]$  &  $\left[\begin{array}{cc} 0 & -\frac{1}{2}\\ \frac{1}{2} &\frac{2}{3}\end{array}\right]$  \\
  \hline
\end{tabular}\label{t1}
\end{footnotesize}
\end{table}

The elements of the matrices entering the Table \ref{t1} are indexed by $x_A$ (rows) and $y_A$ (columns).
Some elements of the utility functions above are negative (loss instead of gain) but this can be easily cured, if necessary, using the symmetry transformations (\ref{a1}). The resulting game may seem slightly complicated but the underlying principles are very simple: (\textit{i}) symmetry with respect to the permutations of players, (\textit{ii}) expressibility of the total payoff $F_A+F_B+F_C$ in terms of Bell operators and (\textit{iii}) saturation of the bound for total payoff following from Bell inequalities. The latter reads in our example
\begin{equation}
F_A+F_B+F_C\leq \frac{9}{4}.\label{a5}
\end{equation}
Actually, in order to obtain the utilities presented in Table \ref{t1}, we have used still one constraint to be discussed below.

The game defined by the utilities given in Table \ref{t1} possesses the correlated Nash equilibria described in Table \ref{t}. The rows in first three columns present the values of $y$'s for $x=0$ and $x=1$.

\begin{table}
\caption{"Pure" Nash equilibria}
\center
\begin{tabular}{|c|c|c|c|c|c|}
\hline
$y_A$ & $y_B$ & $y_C$ & $F_A$ & $F_B$ & $F_C$\\
\hline

$\naw{0,1}$ & $\naw{0,0} $ & $\naw{0,0}$ & $\frac{5}{8}$ & $\frac{13}{16}$ & $\frac{13}{16}$\\

$\naw{0,0}$ & $\naw{0,1} $ & $\naw{0,0}$ & $\frac{13}{16}$ & $\frac{5}{8}$ & $\frac{13}{16}$\\

$\naw{0,0}$ & $\naw{0,0} $ & $\naw{0,1}$ & $\frac{13}{16}$ & $\frac{13}{16}$ & $\frac{5}{8}$\\
\hline
$\naw{1,0}$ & $\naw{0,1} $ & $\naw{0,1}$ & $\frac{11}{8}$ & $\frac{7}{16}$ & $\frac{7}{16}$\\
$\naw{0,1}$ & $\naw{1,0} $ & $\naw{0,1}$ & $\frac{7}{16}$ & $\frac{11}{8}$ & $\frac{7}{16}$\\
$\naw{0,1}$ & $\naw{0,1} $ & $\naw{1,0}$ & $\frac{7}{16}$ & $\frac{7}{16}$ & $\frac{11}{8}$\\
\hline
$\naw{0,1}$ & $\naw{1,1} $ & $\naw{1,1}$ & $\frac{3}{4}$ & $\frac{3}{4}$ & $\frac{3}{4}$\\
$\naw{1,1}$ & $\naw{0,1} $ & $\naw{1,1}$ & $\frac{3}{4}$ & $\frac{3}{4}$ & $\frac{3}{4}$\\
$\naw{1,1}$ & $\naw{1,1} $ & $\naw{0,1}$ & $\frac{3}{4}$ & $\frac{3}{4}$ & $\frac{3}{4}$\\
\hline
\end{tabular}\label{t}
\end{table}
In order to show that the configurations presented in Table \ref{t} provide the Nash equilibria we note first that the relevant probabilities are of the form (\ref{a4}). Consider, for example, the first row in Table \ref{t}. The probabilities corresponding to the strategies entering it read
\begin{equation}
\begin{split}
& p_A\naw{y_A|x_A,\lambda}=\delta_{y_A,x_A}\\
& p_B\naw{y_B|x_B,\lambda}=\delta_{y_B,0}\\
& p_C\naw{y_C|x_C,\lambda}=\delta_{y_C,0}.
\end{split}\label{a6}
\end{equation}

Eqs. (\ref{a6}) define an equilibrium. To see this consider the Alice payoff. Eqs. (\ref{a4}) and (\ref{a6}) yield
\begin{equation}
p\naw{\underline{y}|\underline{x}}=\delta_{y_B,0}\delta_{y_C,0}\int\text{d}\lambda\rho\naw{\lambda}p_A\naw{y_A|x_A,\lambda}\equiv p_A\naw{y_A|x_A}\delta_{y_B,0}\delta_{y_C,0}
\end{equation}
and, consequently,
\begin{equation}
\begin{split}
& F_A=\frac{1}{8}\sum_{\naw{x_A,y_A}}p_A\naw{y_A|x_A}\sum_{x_B,x_C}u_A\naw{x_A,x_B,x_C,y_A,0,0}\equiv\\
& \quad\equiv \frac{1}{8}\sum_{x_A,y_A}p_A\naw{y_A|x_A}u_A\naw{x_A,y_A}.
\end{split}
\end{equation}
$F_A$ should be maximized on the convex set $\sum\limits_{y_A}p_A\naw{y_A|x_A}=1$, $x_A=0,1$; $F_A$ acquires maximum on some of extremal points of this set. The same reasoning applies to Bob and Charlie. So it remains to check the equilibrium condition on $2^6$ strategies $y_i\naw{x_i}$, $i\in\poisson{A,B,C}$. 

By inspecting the Table \ref{t} we see that we have 3 groups, each containing 3 equilibria; each set is invariant under the permutation of players. Two sets represent unfair equilibria; the remaining one contains fair ones. The game is a conflicting interest one as it is clearly seen from Table \ref{t}: there is no common equilibrium preferred by all players. In fact, even if some mixed (i.e. the one with some $0<p\naw{\underline{y}|\underline{x}}<1$) fair equilibrium existed, the payoff of each player could not exceed $\frac{3}{4}$ due to the bound on the total payoff following from Bell inequalities. 

\section{The quantum counterpart of three-player game}

Let us now pass to the quantum case. The density matrix $\rho$ entering eq. (\ref{b2}) is chosen as
\begin{equation}
\rho=\ket{\Psi}\bra{\Psi}
\end{equation}
where $\ket{\Psi}$ is the GHZ state
\begin{equation}
\ket{\Psi}=\frac{1}{\sqrt{2}}\naw{\ket{111}+i\ket{000}}.
\end{equation}
The choice of $\rho$ determines the properties of advisor while the players strategies are described by the probabilities $p\naw{\underline{y}|\underline{x}}$ which, in turn, are determined by choosing the unit vectors $\vec{n}_x^{(A)}$, $\vec{n}_x^{(B)}$ and $\vec{n}_x^{(C)}$; one needs twelve angles to characterize them. This makes the problem complicated. Therefore, we restrict ourselves to the special case $\theta_x^A=\theta_x^B=\theta_x^C=\frac{\pi}{2}$. Let us denote by $\naw{\varphi_1,\varphi_2}$, $\naw{\varphi_3,\varphi_4}$ and $\naw{\varphi_5,\varphi_6}$ the angles characterizing the observables $A_x$, $B_x$ and $C_x$, respectively. It is then not difficult to find the relevant payoffs
\begin{equation}
\begin{split}
& F\equiv F_{A,B,C}=\frac{1}{48}\left( 26+3\sin\naw{\varphi_1+\varphi_3+\varphi_5}+2\sin\naw{\varphi_2+\varphi_3+\varphi_5}+\right.\\
& \quad +2\sin\naw{\varphi_1+\varphi_4+\varphi_5}-3\sin\naw{\varphi_2+\varphi_4+\varphi_5}+2\sin\naw{\varphi_1+\varphi_3+\varphi_6}+\\
& \quad \left.-3\sin\naw{\varphi_2+\varphi_3+\varphi_6}-3\sin\naw{\varphi_1+\varphi_4+\varphi_6}-2\sin\naw{\varphi_2+\varphi_4+\varphi_6}\right).
\end{split}
\end{equation}
We have fixed the values of utility functions in such a way that all payoff functions are equal; this is the additional condition we have mantioned before. Due to this property all Nash equilibria must be fair.

$F$ is invariant under the transformations 
\begin{equation}
\begin{split}
& \varphi_1\rightarrow\varphi_1+\chi_1,\quad \varphi_2\rightarrow\varphi_2+\chi_1,\quad \varphi_3\rightarrow\varphi_3+\chi_2,\quad\varphi_4\rightarrow\varphi_4+\chi_2\\
& \varphi_5\rightarrow\varphi_5+\chi_3,\quad \varphi_6\rightarrow\varphi_6+\chi_3
\end{split}\label{b3}
\end{equation}
provided $\chi_1+\chi_2+\chi_3=2n\pi$. This follows from the relation
\begin{equation}
e^{i\frac{\chi_1}{2}\sigma_3}\otimes e^{i\frac{\chi_2}{2}\sigma_3}\otimes e^{i\frac{\chi_3}{2}\sigma_3}\ket{GHZ}=\naw{-1}^n\ket{GHZ}.
\end{equation}
Maximizing $F$ one obtains the Nash equilibrium. Due to the symmetry (\ref{b3}) we get two parameter family of equilibria. To fix one we put $\varphi_1=\varphi_3=0$. Then the remaining angles (obtained numerically) read $\varphi_2=-\frac{\pi}{2}$, $\varphi_4=-\frac{\pi}{2}$, $\varphi_5=2.1588$, $\varphi_6=0.5880$ (up to the multiples of $2\pi$). The corresponding gain of each player is 
\begin{equation}
F_A=F_B=F_C=0.842
\end{equation} 
We conclude that the quantum version of the game possesses only fair equilibria and the corresponding payoffs are higher than in any classical fair equilibrium which, due to the inequality (\ref{a5}), cannot exceed 0,75. Let us note that our game is genuinely a quantum one (in spite of the restriction $\theta_x^i=\frac{\pi}{2}$ imposed) since the strategies are represented by, in general, noncommuting observables. However, the result obtained (the existence of only fair equilibria) might occur as a consequence of artificial constraint imposed on the $\theta$ angles.
To get some feeling what is going on consider the general quantum game with no constraints on $\theta's$. Let us take into account the unfair equilibrium corresponding to the first row of Table \ref{t}. It is defined by the probabilities $p\naw{\underline{y}|\underline{x}}$ which cannot appear on quantum level. In fact, assume we have six pairs of onedimensional projectors $A_x^y$, $B_x^y$ and $C_x^y$ obeying
\begin{equation}
\bra{\Psi}\naw{A_{x_A}^{y_A}\otimes B_{x_B}^{y_B}\otimes C_{x_C}^{y_C}}\ket{\Psi}=\delta_{y_A,x_A}\delta_{y_B,0}\delta_{y_C,0}.
\end{equation} 
Summing over $y_A$ and $y_B$ yields 
\begin{equation}
\bra{\Psi}\naw{\mathbbm{1}\otimes\mathbbm{1}\otimes C_{x_C}^{y_C}}\ket{\Psi}=\delta_{y_C,0}.
\end{equation}
Now, $\mathbbm{1}\otimes\mathbbm{1}\otimes C_{x_C}^{y_C}$ is a projector so that 
\begin{equation}
\naw{\mathbbm{1}\otimes\mathbbm{1}\otimes C_{x_C}^{y_C}}\ket{\Psi}=\ket{\Psi}
\end{equation}
which is impossible ($\ket{\Psi}\equiv\ket{GHZ}$).

We conclude that not all classical strategies can be reproduced on quantum level. In particular, this concerns strategies leading to unfair equilibria. One can say that quantum entanglement plays here twofold role: it excludes at least some (unfair) equilibria and raises the payoffs corresponding to fair equilibria.

Finally, let us note that our results concern the case of uniform distribution of the player types (cf. eq. (\ref{ca})). If this condition is relaxed new interesting possibilities arise. In the nice recent paper \cite{Auletta} Auletta et al. presented some examples of three-party GHZ games with nonuniform distributions of types; in particular, they constructed a game with the following feature: no no-signalling (superquantum) distribution can help to achieve a better fair equilibrium than that achieved by a quantum strategy.
 However, it should be stressed that the assumption concerning the nonuniform distribution of types is here crucial.

\section{Conclusions}
We have outlined the construction of general three-player game with incomplete information such that: ($i$) it is symmetric under the permutation of players, ($ii$) the upper bound on the total payoff results from Bell inequalities, ($iii$) there exist both fair and unfair Nasha equilibria saturating this bound. Such games are necessarily conflicting interes ones. Although the general formulae are rather involved, the basic assumptions and the algorithm for constructing the game are clearly described which allows to produce easily numerous examples. One example is presented in detail. Contrary to the case of two-player game \cite{Pappa}, \cite{Situ}, \cite{Roy} one has to combine at least two Bell inequalities to obtain a nontrivial game.

A quantum counterpart of the game is obtained by keeping the same utility functions but replacing the classical advisor by a quantum one. As it has been already shown by Pappa et al  \cite{Pappa} the quantum strategies can outperform the classical ones due to the quantum phenomenon of entanglement which leads to the violation of Bell inequalities. The description of entanglement in the three-partite (and multi-partite) case is more complicated than in two-partite one (see, for example, Ref. \cite{Brunner1}). One can consider, for example, the three-partite correlations which are the mixtures of quantum and classical ones \cite{Svetlichny} and construct a three-players game based on Svetlichny inequalities \cite{Situ1}. It is desirable to construct also the three-player games based on Bell inequalities. In such a case one has to use more than one Bell inequality. Another important point which should be mentioned is that in order to ensure the violation of Bell inequalities (which allows the quantum strategies to outperform the classical ones) one has to choose a particular form of quantum advisor. It appears that it can be chosen in such a way that the payoff functions of the players coincide. This happens to be the case in the example considered in Ref. \cite{Pappa} as well as in the one described in the present paper. The quantum game possesses then only fair equilibria.

\subsection*{Acknowledgement}
I would like to thank Prof. Piotr Kosi\'nski  for helpful discussion and useful remarks. The research was supported by the NCN Grant no. DEC-2012/05/D/ST2/00754.

\end{document}